\documentclass[aps,twocolumn,groupedaddress,showpacs]{revtex4}
\usepackage{amsmath,amscd,amssymb,epsfig}
\begin{document}
\bibliographystyle{apsrev}

\title[]{Anomalous escape
governed by thermal $1/f$ noise}
\author{I. Goychuk and P. H\"anggi}
\author{}
\affiliation{University of Augsburg, Institute of Physics,
Universit\"atsstr. 1, D-86135 Augsburg, Germany}

\date{\today}

\begin{abstract}

We present an analytic study  for subdiffusive escape  of overdamped
particles out of a cusp-shaped parabolic potential well which are
driven by thermal, fractional Gaussian noise with a
$1/\omega^{1-\alpha}$ power spectrum. This long-standing challenge
becomes mathematically tractable by use of  a generalized Langevin
dynamics via its corresponding non-Markovian, time-convolutionless
master equation: We find that the escape  is governed asymptotically
by a power law whose exponent depends {\it exponentially} on the
ratio of barrier height and temperature. This result is in distinct
contrast to a description with a corresponding subdiffusive
fractional Fokker-Planck approach; thus providing experimentalists
an amenable testbed to differentiate between the two escape
scenarios.

\end{abstract}


\maketitle

The theme of anomalous sub-diffusion and rate kinetics continuous to
flourish over the last years. This topic is driven by the
availability of a wealth of intriguing experimental data, ranging
from anomalous diffusion in amorphous materials, quantum dots,
protein dynamics, actin networks, and biological cells
\cite{Scher,Austin,metzler2000,Sokolov,TangMarcus,
Yang,Xie2,actin,Caspi,cell,PRE04}.  Suitable theoretical
descriptions derive from continuous time random walks (CTRW)
\cite{Scher,Weiss}, the   CTRW-based fractional Fokker-Planck
 (FFP)-approach \cite{MetzlerPRL,PRE06}, or the generalized
Langevin equation (GLE) \cite{HTB90,Bao2006}. 
The GLE-subdiffusion implies
power-law-correlated thermal forces (or fractional Gaussian noise
(fGn) \cite{Mandelbrot}) possessing  infinite memory  with a
$1/\omega^{1-\alpha}$ power spectrum \cite{Kupferman,Xie2}. Such
random forces emerge when coupling the system to sub-Ohmic thermal
baths with spectral densities $J(\omega)\propto \omega^{\alpha},\;
0<\alpha<1$ \cite{Bao2006}.

Recent experiments on anomalous conformational subdiffusion in
electron-transferring proteins have been successfully modeled within
a FFP equation \cite{Yang}.   Soon after, however, the fGn-Langevin
approach has been shown to describe the experimental data even more
convincingly \cite{Xie2}. Both approaches are consistent with
molecular dynamics simulations \cite{moldyn}. In addition, both
schemes are consistent with the laws of thermodynamics. Fundamental
differences manifest themselves, however, when one considers the
escape dynamics. The description of  fGn-driven escape
presents a long-standing, timely problem. This is so because this
complexity, in contrast to the  Kramers escape dynamics
\cite{HTB90}, then generally no longer allows for a detailed, even
approximate solution. Recent attempts to solve this challenge can be
found in Refs. \cite{new}, although not yielding
a proper solution. Thus, the issue of 1/f-noise driven
escape remains an intriguing  open problem which continues to haunt
the literature.

Our  analytic work  deals with the unique, but exactly solvable
challenge of an escape driven by 1/f-noise out of a cusp-shaped
parabolic potential. In doing so we demonstrate that the resulting
escape dynamics is scale-free, i.e. it is governed by a power law.
This in turn invalidates a rate description. Moreover, the escape
dynamics within the fGn-Langevin description is {\it exponentially
sensitive to temperature}. This result is in marked contrast to the
description within a FFP-dynamics which instead yields a {\it
temperature-independent} power law exponent. Thus, in contrast to
the finite mean first passage time (MFPT) result  in \cite{new} our
main result exhibits an infinite MFPT, being consistent with a
strict subdiffusive escape dynamics. A
rate description  emerges only when invoking a physically plausible low
frequency regularization of the noise spectrum.

\begin{figure}[ht]
 \includegraphics[width=7.5cm]{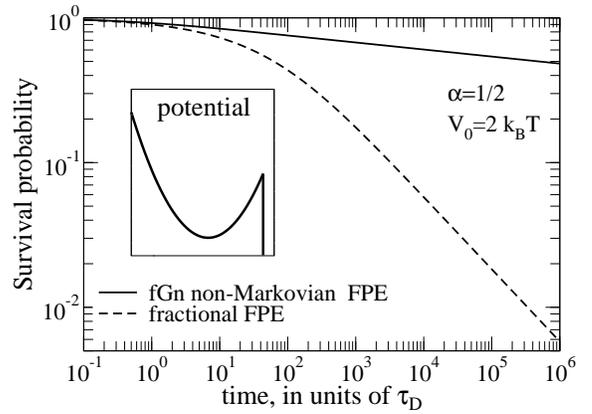}
 \caption{\label{Fig1}
 Comparison between the escape dynamics described by the fractional
 Fokker-Planck equation and the overdamped, non-Markovian GLE-dynamics 
 in (\ref{Langevin})-(\ref{fBm})
 in a parabolic cusp potential, depicted with the inset.
 }
 \end{figure}

\textit{GLE-approach.} We start out from the GLE for a particle of
mass $m$ moving in the potential $V(x)$ \cite{HTB90}; i.e.,
\begin{eqnarray}\label{Langevin}
m\ddot x +\int_{0}^{t}\eta(t-t')\dot x(t')dt'+V'(x)=\xi(t).
\end{eqnarray}
The autocorrelation function $\langle \xi(t)\xi(t')\rangle$ of
thermal Gaussian noise $\xi(t)$ and the frictional kernel $\eta(t)$
are related by the usual fluctuation-dissipation relation
\cite{HTB90}:
\begin{eqnarray}\label{FDR}
\langle \xi(t)\xi(t')\rangle =  k_B T\eta(|t-t'|).
\end{eqnarray}
In the following we consider the overdamped limit with $m\to 0$;
i.e. the  velocity is thermally relaxed at each instant of time.
Moreover, we assume that the particles are initially localized in a
metastable parabolic well  at $x_0$, cf. the inset in Fig.
(\ref{Fig1}). The starting probability density then is $P(x,
t=0)=\delta(x-x_0)$. The corresponding non-Markovian master equation
for $P(x,t)$ for this GLE is generally not known: For arbitrary
physical memory-friction $\eta(t)$ this task is known for two  cases
only; namely (i) a linear potential $V(x)=-F_0\;x$, including the
case of free diffusion, i.e. $F_0=0$ and (ii) a parabolic potential
$V(x)=\kappa x^2/2$. The procedure to obtain the master equation is
well known: It is solely rooted in the Gaussian nature of $x(t)$
 \cite{Adelman,HanggiOld,HanggiMojtabai,Hynes,Kupferman}.
The result is a time-convolutionless master equation for $P(x,t)$
obeying a Fokker-Planck form  with a time-dependent diffusion
coefficient $D(t)$ \cite{HanggiOld,Hynes}, reading
\begin{eqnarray}\label{Smoluch}
\frac{\partial P(x,t)}{\partial t}=D(t)
\frac{\partial}{\partial x}\left (e^{-\beta V(x)}
\frac{\partial}{\partial x}e^{\beta V(x)} P(x,t)  \right).
\end{eqnarray}
Here,  $\beta=1/(k_BT)$. Notably, $D(t)$ does not depend on $x_0$,
but is dependent on $V(x)$ and memory friction $\eta(t)$.
For a quadratic potential,
it can be expressed via the relaxation function $\theta(t)$ of
position fluctuations as \cite{HanggiOld, Hynes}
\begin{eqnarray}\label{Dquad}
D(t)=-l_T^2\frac{d}{dt}\ln \theta(t)\;,
\end{eqnarray}
where $l_T=\sqrt{k_BT/\kappa}$ is the length scale of thermal
fluctuations. The Laplace-transform of $\theta(t)$ is related to the
memory friction $\eta(t)$ by $\tilde \theta(s)=\tilde
\eta(s)/[\kappa+s\tilde \eta(s)]\;$.

We next use a power-law friction kernel $\eta(t)$, reading
\begin{eqnarray}\label{fBm}
\eta(t)=\frac{\eta_{\alpha}}{\Gamma(1-\alpha)}
\frac{1}{|t|^{\alpha}}\;, \; 0<\alpha<1 \;.
\end{eqnarray}
This friction $\eta(t)$ yields an anomalous, free ($V(x) = 0$)
subdiffusion with $\langle\delta x^2(t)\rangle  =2
K_{\alpha}t^{\alpha}/\Gamma(1+\alpha)$, where the  anomalous
diffusion coefficient $K_{\alpha}=k_BT/\eta_{\alpha}$ obeys a
generalized Einstein relation. For a parabolic potential this yields
the relaxation function
\begin{eqnarray}\label{ML}
\theta(t)=E_{\alpha}[-(t/\tau_D)^{\alpha}] \;,
\end{eqnarray}
where $\tau_D=(\eta_{\alpha}/\kappa)^{1/\alpha}$, and
$E_{\alpha}(z)$ is the Mittag-Leffler function, i.e.,
$E_{\alpha}(z)=\sum_{n=0}^{\infty}z^n/\Gamma (\alpha n+1)$
\cite{GorenfloMainardi}. It corresponds to the Cole-Cole model of
glassy dielectric media \cite{Hilfer}, whereas the limit $\alpha\to
1$ corresponds to  an exponential relaxation with $E_1(x)=\exp (x)$.

The thermal fGn $\xi(t)$ is the time derivative of  fractional
Brownian motion (fBm) \cite{Mandelbrot} with a power spectrum, $
S_{\xi}(\omega)=2k_B
T\eta_{\alpha}\sin(\pi\alpha/2)/\omega^{1-\alpha}$. For this thermal
fGn  the  GLE in (\ref{Langevin}) with $m=0$ can formally
identically be recast as  the "fractional" Langevin equation, i.e., 
$\eta_{\alpha}D^{\alpha}_* x(t)+V'(x)=\xi(t)$,
wherein $ D_{*}^{\alpha} x (t) = (1/\Gamma(1 - \alpha)) \int_0^t
\mathrm{d} t' (t - t')^{-\alpha} \dot x(t')$ is the operator of the
fractional Caputo derivative \cite{GorenfloMainardi}.

\textit{FFP-approach.} Alternatively, if instead of the fGn dwelling
in a potential in (\ref{Langevin}), (\ref{FDR}), (\ref{fBm}) we  use  a
modeling in terms of an overdamped, fractional Fokker-Planck
equation description \cite{MetzlerPRL,PRE06}, the probability
density obeys
\begin{eqnarray}\label{FFPE}
D_{*}^{\alpha} P(x,t)=K_{\alpha} \frac{\partial}{\partial x}\left
(e^{-\beta V(x)} \frac{\partial}{\partial x}e^{\beta V(x)} P(x,t)
\right)\;.
\end{eqnarray}
This result derives from an underlying continuous time random walk
description of subdiffusion \cite{Scher}. It as well has an
associated Langevin equation in a random operational time $t(\tau)$
\cite{Stanislavsky} which, however, is profoundly {\it different}
from the  GLE.

\textit{Subdiffusive dynamics  dwelling in a parabolic potential.}
The time-convolutionless master equation (\ref{Smoluch}) of the GLE
in (\ref{Langevin}), (\ref{FDR}), (\ref{fBm}) 
can be solved exactly for a parabolic
potential \cite{remark2}: We first transform $P(x,t)$  as
$P(x,t)=\exp(-\beta V(x)/2) W(x,t)$ and separate the variables,
$W(x,t)=Y(x)\Phi(t)$. For the coordinate-dependent part this yields
a spectral representation, reading
\begin{eqnarray}\label{Y}
Y^{''}_n(x)+\frac{\beta}{2}\kappa\left (1-\frac{\beta}{2}\kappa x^2
+ 2 \lambda_n/(\beta\kappa) \right)
Y_n(x)=0,
\end{eqnarray}
where $\lambda_n$ and $Y_n(x)$ are the corresponding spectral
eigenvalues and eigenfunctions. The  functions  $\Phi_n(t)$ obey:
\begin{eqnarray}\label{relax1}
\dot \Phi_n(t)=-\lambda_n D(t) \Phi_n(t).
\end{eqnarray}
By use of (\ref{Dquad}) the exact solutions of (\ref{relax1}) read
\begin{eqnarray}
\Phi_n(t)=[\theta(t)]^{s_n},
\end{eqnarray}
where $s_n:=l_T^2\lambda_n$. These findings  yield for the explicit
solution for the probability density $P(x,t)$ the result,
\begin{eqnarray}\label{exact1}
P(x,t)=\exp(-\beta\kappa x^2/4)
\sum_n c_n Y_n(x)[\theta(t)]^{s_n},
\end{eqnarray}
where the expansion coefficients $c_n$ are determined from the
initial probability density $P(x,t=0)$. The spectrum reads $s_n=n$,
$n=0,1,2,...$. Moreover, the functions $Y_n(x)$ are  given in terms
of Hermite functions \cite{HTB90}. The dynamics of the probability
evolution is thus ruled by the relaxation function $\theta(t)$ in
Eq. (\ref{ML}). Remarkably, the relaxation of the mean value
$\langle x(t)\rangle$ follows precisely to the same law as in the
case of a FFP description \cite{MetzlerPRL}. This is surprising
because the general solution $P(x,t)$ differs markedly from that of
the  FFP equation in \cite{MetzlerPRL}. The solution in the latter
case is obtained from (\ref{exact1}) by substituting  therein
$E_{\alpha}[-n(t/\tau_D)^\alpha]$ for
$(E_{\alpha}[-(t/\tau_D)^\alpha])^n$ within $\theta^n(t)$. This
follows from the  fact that for the FFP in (\ref{FFPE}), equation
(\ref{relax1}) is replaced  by $
D_{*}^{\alpha} \Phi_n(t)=-\lambda_n K_{\alpha} \Phi_n(t)$ 
\cite{MetzlerPRL}.

Because the integral over $\theta(t)$ in Eq. (\ref{ML}) is not
finite, the statistical mean of random escape times is subject to a
divergence, thus invalidating a rate description. Because inter-well
transitions can always be broken up into the two steps, i.e. (i)
reaching the barrier region from  well bottom and (ii) a barrier
(re)-crossing to an adjacent well, this implies as well a diverging
MFPT  when considering  more generic situations as the stylized one
addressed next.

\textit{Escape out of parabolic cusp potential.} Let us impose next
an infinitely sharp potential cutoff at $x_a=L\gg l_T$, see the
inset in Fig. \ref{Fig1}. This sharp cut-off is identical to an
absorbing boundary condition, satisfying $P(x,t)= 0 $ for $x \ge L$.
The Gaussian approximation for the GLE therefore remains valid
inside the parabolic cusp potential \cite{newremark}. The solution
of eq. (\ref{Smoluch}) for the corresponding boundary value problem
now reads anew (for $-\infty <x\leq L$):
\begin{eqnarray}
Y_n(x)=U\left (-s_n-\frac{1}{2},-x/l_T \right) \;,
\end{eqnarray}
where $U(\nu,x)$ denotes the parabolic cylinder function. The
spectrum for this case reads $\lambda_n =s_n/l_T^2$, being
determined by the solutions of transcendental equation
\begin{eqnarray}\label{spectrum}
U\left (-s_n-\frac{1}{2},-L/l_T\right)=0.
\end{eqnarray}
For $L\to \infty$, $s_n$ approaches again $n$. For large $L\gg l_T$
the decay of the survival probability inside the well $P_{\rm
SP}(t)=\int_{-\infty}^L P_{\rm cusp}(x,t)dx$ is ruled by the lowest
eigenvalue; i.e.,
\begin{eqnarray}\label{fin}
P_{\rm SP}(t)\approx [\theta(t)]^{s_0}.
\end{eqnarray}
This constitutes our first central result.  The value $s_0$ is given
by the numerical solution of Eq. (\ref{spectrum}). Remarkably, it is
well approximated by the inverse of the properly scaled  mean first
passage time  of the corresponding,  memoryless Markovian problem
yielding, cf. in Ref. \cite{HTB90}:
\begin{eqnarray}\label{Mark}
s_0^{-1}=l_T^{-2}\int_0^{L}dy\int_{-\infty}^y dx\exp(\beta[V(y)-V(x)]).
\end{eqnarray}
This yields $ s_0=F\left (\frac{V_0}{k_BT}\right)$, where
$V_0=\kappa L^2/2$ is the barrier height and
$1/F(z)=\sqrt{\pi}\int_{0}^{\sqrt{z}}e^{y^2}[1+{\rm erf}(y)]dy$. For
example, for $L=2l_T$ the exact value of $s_0$ is
$s_0^{\rm (exact)}=0.09727$ while (\ref{Mark}) yields
$s_0^{\rm (approx.)}=0.09589$. The difference is already less than
1.5$\%$ and  rapidly  diminishes with increasing $L$. As a main
trend, $s_0$ decreases approximatively exponentially $\propto
\exp(-V_0/k_BT)$, thus displaying a typical Arrhenius dependence.

Within the approximation of  (\ref{fin}) the MFPT of the
non-Markovian escape dynamics is given by:
\begin{eqnarray}
\langle \tau\rangle =\int_0^{\infty}P_{\rm SP}(t)dt =
\int_0^{\infty}[\theta(t)]^{s_0}dt \;,
\end{eqnarray}
being indeed  very distinct  from the Markovian case. For the
fGn-GLE model, denoted in the following by $P_{\rm SP}^{\rm GLE}$, we find
from (\ref{fin}) $P_{\rm SP}^{\rm GLE}(t)\approx \left (E_{\alpha}\left
[-(t/\tau_D)^{\alpha} \right ] \right )^{s_0}$, and thus the MFPT
diverges, i.e. $\langle \tau\rangle=\infty$.

Likewise, the high-barrier solution of the FFP equation in
(\ref{FFPE}) is given by $P_{\rm SP}^{\rm FFP}(t)\approx
E_{\alpha}\left [-s_0(t/\tau_D)^{\alpha}\right]$, yielding again no
finite value for the MFPT. The asymptotic long-time behaviors 
differ distinctly  in these two models:
\begin{eqnarray}\label{fin5}
P_{\mathrm{ SP}}^{\mathrm{ GLE}}(t)& \sim & \Gamma(1-\alpha)^{-s_0}
\left( \frac{\tau_D}{t}\right)^{s_0\alpha},\\
P_{\mathrm{ SP}}^{\mathrm{ FFP}}(t)& \sim & \frac{1}{s_0
\Gamma(1-\alpha)} \left (\frac{\tau_D}{t}\right)^{\alpha}\;,
\end{eqnarray}
where $\tau_D=(\eta_{\alpha}/\kappa)^{1/\alpha}$. In particular, for
the fGn-GLE model, the power law exponent $s_0\alpha$ depends
exponentially    on the barrier height and the (inverse)
temperature. In contrast, for  the FFP-theory this power law
exponent just equals  the subdiffusive power law exponent $\alpha$.


\begin{figure}[ht]
\vspace{0.8cm}
 \includegraphics[width=7.5cm]{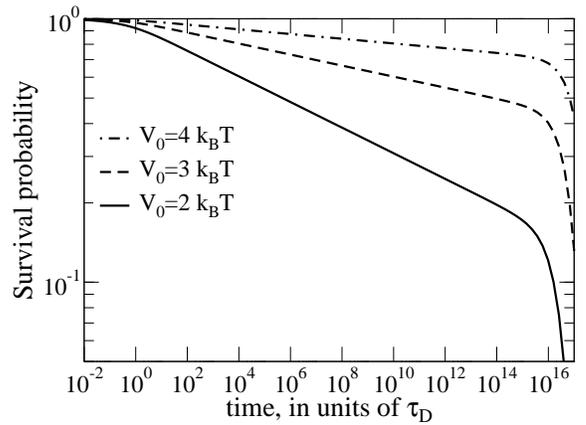}
 \caption{\label{Fig2}
 Survival probability for  a modified fGn-GLE model using a regularization.
 An exponential relaxation tail emerges for $t>\omega_c^{-1}$.
The chosen parameters are: $\tau_D \sim {\rm 1\;ps}$, $\omega_c \sim
{\rm 1\; hour^{-1}}$ and $\alpha=1/2$.
 }
 \end{figure}
 \vspace{-0.1cm}
 Fig. \ref{Fig1} depicts a comparison between two escape dynamics for
$\alpha=1/2$ \cite{Xie2}, where $\theta(t)=\exp(t/\tau_D){\rm
erfc}(\sqrt{t/\tau_D})$, and for $\beta V_0=2$. The  FFP-escape
dynamics overall proceeds faster. The initial kinetic stages are
identical. A detection of a power-law   escape that is exponentially
sensitive to temperature would corroborate the GLE based approach;
while a temperature-{\it independent} power law decay would favor
the FFP-approach.

\textit{Role of memory cut-off}. Both  considered theoretical models
contain a physical drawback: Random forces obeying  a true
$1/\omega^{1-\alpha}$ feature in the power spectrum are not
physical; i.e.  a low-frequency regularization must always emerge on
physical grounds, (implying that $S(\omega=0)$ is finite)
\cite{Weissman}. To account for this physical requirement  we introduce
an exponential cutoff $\exp(-\omega_c t)$ with a  small frequency
$\omega_c$ for the memory kernel (\ref{fBm}), yielding $S(\omega=0)=
2k_B T\eta_{\alpha}/\omega_c^{1-\alpha}$.
The corresponding relaxation function $\theta(t)$ now exhibits  an
exponential decay $\theta(t)\propto \exp(-\omega_c t)$, for $t\geq
1/\omega_c$. The memory kernel $\eta(t)$ becomes integrable so that
the MFPT $\langle \tau\rangle$  exists. As a consequence, a
non-Markovian rate description now becomes valid \cite{HTB90}. In
practice, however, a feasible rate description fails whenever the
main part of the escape dynamics occurs  within a distinct
non-exponential, power law regime which extends over many temporal
decades. This feature is elucidated with Fig. \ref{Fig2}. A valid
rate description, although with an extremely small rate is restored
by either lowering the temperature, or likewise, by increasing the
barrier height. The top curve (highest barrier case) in Fig.
\ref{Fig2}, depicts  this trend. Our results corroborate also with
the numerical simulations of bistable dynamics \cite{Min}, where a
numerical cut-off is intrinsically present. Using a memory cut-off
within the CTRW description for the FFP in Eq. (\ref{FFPE}) does
result as well in an asymptotically finite rate. The intermediate
power law will exhibit, however, also no distinct temperature
dependence, being again in a clear contrast with the subdiffusive
GLE description.

In conclusion,  we put forward an analytical treatment of the
survival probability for  the non-Markovian escape from a
cusp-shaped well when anomalous subdiffusion is acting. Then, the
MFPT diverges which in turn invalidates a rate description. The
sensible physical requirement of a low-frequency regularization
enables one to restore a rate theory description that is valid for
sufficiently high barriers, or very low temperatures. The
single-molecular enzyme kinetics \cite{Yang,Xie2} might present a
suitable candidate to validate experimentally the intriguing
crossover between an exponential and a power law kinetic regime
which crucially depends on temperature.

This work was supported by the
German Excellence Initiative via the \textit {Nanosystems Initiative
Munich (NIM)}.


\begin{references}

\bibitem{Scher}
H. Scher, E.W. Montroll, Phys. Rev. E {\bf 12}, 2455
(1975); M. Shlesinger, J. Stat. Phys. {\bf 10}, 421 (1974).

\bibitem{Austin}R. H. Austin, K. W. Beeson, L. Eisenstein,
H. Frauenfelder, and I.C. Gunsalus, Biochemistry {\bf 14}, 5355
(1975).

\bibitem{metzler2000} R. Metzler, J. Klafter, Phys. Rep.
\textbf{339}, 1 (2000).
%
\bibitem{Sokolov} I.M. Sokolov, J. Klafter, A. Blumen,
Phys. Today \textbf{55} (11), 48 (2002); J. Klafter, I.M. Sokolov,
Physics World, August, 29 (2005).

\bibitem{TangMarcus}J. Tang and R.A. Marcus, Phys. Rev. Lett.
{\bf 95}, 107401 (2005).

\bibitem{Yang} H. Yang \textit{et al.},
Science \textbf{302}, 262 (2003); H. Yang and
X. S. Xie, J. Chem. Phys. {\bf 117}, 10965 (2002).

\bibitem{Xie2}S.C. Kou, X.S. Xie,
Phys. Rev. Lett. \textbf{93}, 180603 (2004);
W. Min, G. Luo, B. J. Cherayil, S. C. Kou, and X. S. Xie,
Phys. Rev. Lett. \textbf{94}, 198302 (2005).

\bibitem{actin}F. Amblard, A. C. Maggs, B. Yurke, A. N. Pargellis,
and S. Leibler, Phys. Rev. Lett. {\bf 77} 4470
(1996);
 I.Y. Wong {\it et al.}, Phys. Rev. Lett. {\bf 92},
178101 (2004).

\bibitem{Caspi}A. Caspi, R. Granek, and M. Elbaum, Phys. Rev. E {\bf 66},
011916 (2002).

\bibitem{cell}M. J. Saxton, Biophys. J. {\bf 81}, 2226 (2001);
I. M. Toli\'c-N\o rrelykke, E.-L. Munteanu, G. Thon,
L. Oddershede, and  K. Berg-S\o rensen, Phys. Rev. Lett. {\bf 93},
078102 (2004); M. Weiss, M. Elsner, F. Kartberg, and T. Nilsson,
Biophys. J. {\bf 87}, 3518 (2004); I. Golding and E.C. Cox, Phys. Rev. Lett.
{\bf 96}, 098102 (2006).

\bibitem{PRE04}I. Goychuk and P. H\"anggi, Phys. Rev. E {\bf 70},
051915 (2004).

\bibitem{Weiss} G.H. Weiss, \textit{Aspects and Applications
of the Random Walk} (North-Holland, Amsterdam, 1994).

\bibitem{MetzlerPRL}R. Metzler, E. Barkai, J. Klafter,
Phys. Rev. Lett. {\bf 82}, 3563 (1999); R. Metzler, E. Barkai, J. Klafter,
Europhys. Lett. {\bf 46}, 431 (1999); E. Barkai, Phys. Rev. E {\bf
63}, 46118 (2001).

\bibitem{PRE06}I. Goychuk, E. Heinsalu, M.
Patriarca, G. Schmid, and P. H\"anggi, Phys. Rev. E {\bf 73},
020101(R) (2006).

\bibitem{HTB90} P. H\"anggi, P. Talkner, M. Borkovec,
Rev. Mod. Phys. \textbf{62}, 251 (1990).


\bibitem{Bao2006} J.~D.~Bao, Y. Z. Zhuo, F. A. Oliveira,
and P. H\"anggi, Phys.~Rev.~ E \textbf{74},
061111 (2006).

\bibitem{Mandelbrot}B. B. Mandelbrot and J. W. van Ness, SIAM Rev.
{\bf 10}, 422 (1968).

\bibitem{Kupferman}
K.G. Wang and M. Tokuyama, Physica A {265}, 341 (1999);
R. Kupferman, J. Stat. Phys. {\bf 114}, 291 (2004).



\bibitem{moldyn} A. R. Bizzarri and S. Cannistraro, J. Phys. Chem. B
{\bf 106}, 6617 (2002);
G. R. Kneller and K. Hinsen, J. Chem. Phys. {\bf 121},
10278 (2004);  G. Luo, I. Andricioaei, X. S. Xie, and
M. Karplus, J. Phys. Chem. B {\bf 110}, 9363 (2006).

\bibitem{new}S. Chaudhury and B. J. Cherayil, J. Chem. Phys. {\bf 125},
024904 (2006); ibid. {\bf 125}, 114166 (2006).

\bibitem{Adelman} S. A. Adelman, J. Chem. Phys. {\bf 64}, 124 (1976).

\bibitem{HanggiOld}P. H\"anggi and H. Thomas, Z. Physik B {\bf 26},
85 (1977); P. H\"anggi, H. Thomas, H. Grabert, and P. Talkner,
J. Stat. Phys. {\bf 18}, 155 (1978); P. H\"anggi, Z. Physik B
{\bf 31}, 407 (1978).


\bibitem{HanggiMojtabai} P. H\"anggi and F. Mojtabai,
Phys. Rev. A {\bf 26}, 1168 (1982).

\bibitem{Hynes}J. T. Hynes, J. Phys. Chem. {\bf 90}, 3701 (1986).


\bibitem{Weissman}M. B. Weissman, Rev. Mod. Phys. {\bf 60}, 537 (1988).

\bibitem{GorenfloMainardi} R. Gorenflo, F. Mainardi, in:
\textit{Fractals and Fractional Calculus in Continuum Mechanics},
edited by A. Carpinteri, F. Mainardi (Springer, Wien, 1997), pp.
223-276.

\bibitem{Hilfer}K. Weron, M. Kotulski, Physica A {\bf 232}, 180 (1996);
R. Hilfer, J. Non-Cryst. Sol. {\bf 305}, 122 (2002).

\bibitem{Stanislavsky} A.~A.~Stanislavsky, Phys.~Rev.~E {\bf 67},
021111 (2003).

\bibitem{remark2} Even if the exact
Green function $P(x,t|x_0,0)=\exp[-(x-x_0\theta(t))^2/(2l_T^2
(1-\theta^2(t))]/\sqrt{2\pi l_T^2 [1-\theta^2(t)]}$ is known
\cite{Hynes}, the resulting eigenfunctions expansion is required
nevertheless for the escape problem.

\bibitem{newremark}
Here, a non-natural boundary condition must be used. Our case of a
cusp-shaped parabolic well with its a sudden drop to $-\infty$
mimics an absorbing line, $x\ge L \gg l_T$. Any initial distribution
relaxes on the time scale $\theta(t)$ to a quasi-equilibrium
Gaussian density with a width around $l_T$, and gradually decays due
to escape.

\bibitem{Min}W. Min and X. S. Xie, Phys. Rev. E {\bf 73},
010902(R) (2006).

\end{references}
\end{document}